\documentclass[aps,reprint,twocolumn,a4paper,superscriptaddress]{revtex4-1}

\usepackage{graphicx}
\usepackage{dcolumn}
\usepackage{bm}
\usepackage{setspace}
\usepackage{amsmath}
\usepackage{amssymb}

\begin{document}

\singlespacing

\title{Anomalous behavior of the $E_u^1$ infrared-active phonon mode in a Bi$_{2-x}$Sr$_x$Se$_3$ crystal}

\author{A. A. Melnikov}
\email{melnikov@isan.troitsk.ru}
\affiliation {Institute for Spectroscopy RAS, Fizicheskaya 5, Troitsk, Moscow, 108840 Russia}
\author{K. N. Boldyrev}
\affiliation {Institute for Spectroscopy RAS, Fizicheskaya 5, Troitsk, Moscow, 108840 Russia}
\author{Yu. G. Selivanov}
\affiliation {P. N. Lebedev Physical Institute RAS, Moscow, 119991 Russia}
\author{S. V. Chekalin}
\affiliation {Institute for Spectroscopy RAS, Fizicheskaya 5, Troitsk, Moscow, 108840 Russia}

\begin{abstract}

We have studied spectral evolution of the $E_u^1$ phonon line of a topological insulator Bi$_{2-x}$Sr$_x$Se$_3$  with temperature. Unlike the Raman-active phonons, the $E_u^1$ mode demonstrates softening upon cooling the crystal, and the corresponding spectral line acquires a pronounced Fano-like shape at temperatures $T\lesssim100$ K. We interpret the latter effect as a signature of specific coupling of the bulk infrared active phonons to surface Dirac electrons. Using coherent resonant excitation of the $E_u^1$ mode as a surface sensitive tool, we have detected softening of the surface counterpart of the bulk $E_u^1$ phonon mode upon strontium doping. This observation can be an evidence of enhanced electron-phonon interaction at the surface of the Bi$_{2-x}$Sr$_x$Se$_3$ crystal.

\end{abstract}

\maketitle

Bi$_2$Se$_3$, as well as related compounds from the family of bismuth and antimony chalcogenides, has unique electronic properties. Common to 3D topological insulators, surface electronic states of these crystals are characterized by linear Dirac-like dispersion with spin helicity \cite{Hasan, Qi}. The spin-momentum locking protects surface charge carriers from backscattering on nonmagnetic impurities and enables spin-polarized surface currents. Therefore, such crystals are prospective for potential applications in spintronics and quantum computation \cite{Pesin, He}. Special attention is given to the problem of superconductivity in Bi$_2$Se$_3$. Since it was discovered that introduction of Cu atoms into tetradymite structure results in a superconducting transition at accessible temperatures \cite{Hor}, the crystals of Bi$_2$Se$_3$ doped with Cu, Sr, or Nb atoms are actively studied as potential topological superconductors --- exotic solids with a superconducting gap in the bulk and gapless Majorana states on the surface \cite{Ando, Valkov, Kuntsevich1, Kulbachinskii}. Moreover, a number of experiments are evidence in favor of electronic nematic ordering upon the superconducting transition of electron-doped Bi$_2$Se$_3$ crystals \cite{Fu, Yonezawa}. 

A key physical process for a superconductor is the electron-phonon interaction. Its character in the superconducting state of electron-doped Bi$_2$Se$_3$ crystals and the role of lattice in the supposed electronic nematic transition are still unclear and are hotly debated. One of the important questions is how rather low electronic concentrations (e. g. $\sim 10^{19}$ cm$^{-3}$ for Sr-doped Bi$_2$Se$_3$) can lead to temperatures of the superconducting transition as large as several kelvins \cite{Almoalem}. For such ``dilute'' superconductors various unconventional pairing mechanisms are currently discussed, including coupling via optical phonons, with a particular emphasis on transverse infrared-active modes \cite{Marel, Sadovskii}. 

In the present work we focus on the properties of the interaction between electrons and infrared-active TO phonons in Bi$_{2-x}$Sr$_x$Se$_3$ accessible using methods of far-infrared spectroscopy. A natural consequence of the interaction of an optical phonon mode with electrons in a solid is the characteristic Fano profile of the corresponding line in the spectrum of light reflected or inelastically scattered from the sample \cite{Fano, Olego, Ager, Brown, Tang, Xu}. In this case vibrational levels play the role of discrete states of the Fano model, while the continuum can be represented by electronic interband transitions in narrow band gap semiconductors or semimetals, intraband transitions, assisted by defects and impurities, or strongly broadened discrete transitions (e. g. between impurity states).

The transverse optical phonon mode of Bi$_2$Se$_3$ that has $E_u^1$ symmetry is readily observable in reflectivity spectra of bulk crystals and thin films. In a number of experimental studies the Fano shape of the corresponding line was detected \cite{LaForge, Dordevic1, Post, Sim, Park, Dordevic2, Mallett}, while some authors used magnetic field \cite{LaForge, Dordevic2} and band engineering \cite{Sim} to control the Fano asymmetry parameter. The effect was ascribed to the interaction of $E_u^1$ phonons with topological electronic surface states, with plasmons, or with electrons in general, without specifying the mechanism. At the same time there exists a certain ambiguity of the experimental results obtained so far, and the reported magnitude, sign, and temperature dependence of the Fano effect measured for the $E_u^1$ phonon line in Bi$_2$Se$_3$ vary considerably.

In the present work we show that in high quality epitaxial single-crystal Bi$_2$Se$_3$ films the asymmetry of the $E_u^1$ line is small at room temperature and vanishes upon cooling. In contrast, for strontium-doped Bi$_2$Se$_3$ this spectral line demonstrates a pronounced Fano profile at temperatures below $\sim$ 100 K. In addition, we present an evidence of softening of the surface counterpart of the $E_u^1$ phonon mode in Bi$_{2-x}$Sr$_x$Se$_3$. We discuss these results in the context of interaction of surface electronic states of binary and strontium-doped Bi$_2$Se$_3$ with infrared-active $E_u^1$ phonons. 

The samples that were studied in our experiments were heterostructures that consisted of crystalline BaF$_2$ substrates, on the surface of which $\sim$ 30 nm thick films of Bi$_2$Se$_3$ or Bi$_{2-x}$Sr$_x$Se$_3$ (x = 6.2 \%) were grown using molecular beam epitaxy. The films were \textit{in situ} protected from ambient air by a thin layer of BaF$_2$ (30--40 nm). The growth procedures were described elsewhere along with specific original methods that allowed the production of epitaxial films of high crystalline perfection \cite{Volosheniuk, Kuntsevich, Oveshnikov}. Stationary terahertz reflectance spectra were measured using the vacuum Fourier spectrometer Bruker IFS 125 HR equipped with an optical helium closed cycle cryostat. In order to obtain additional spectral information on the phonons we used the method of coherent lattice excitation by a powerful terahertz pulse that was described in detail in our previous works \cite{Melnikov1, Melnikov2}. Briefly, a picosecond single-cycle terahertz pulse generates coherent oscillations of atoms of a crystal having a certain symmetry corresponding to particular phonon modes of this crystal. IR-active modes are excited directly, while Raman-active modes -- via a specific nonlinear light scattering mechanism (likely to be the so-called sum frequency Raman scattering \cite{Maehrlein}). A weak delayed femtosecond laser pulse is used to detect the temporal evolution of the anisotropy of the refractive index of the sample created by the pump terahertz pulse. Coherent phonons appear in the recorded signal as damped oscillations at the frequencies that correspond to the excited modes. Performing numerical Fourier transform of the signal allows one to obtain spectral lines of these phonon modes. As we have shown in our previous works, using this method it is possible to detect several bulk Raman-active phonon modes of Bi$_2$Se$_3$ and the surface counterpart of the bulk $E_u^1$ mode, which modulates polarizability of the topmost layers of the crystal due to symmetry breaking at the surface. In comparison to the standard Raman scattering technique this approach allows minimization of lattice heating and excitation of electrons of the sample and in certain cases adds surface selectivity.

\begin{figure}
\begin{center}
\includegraphics{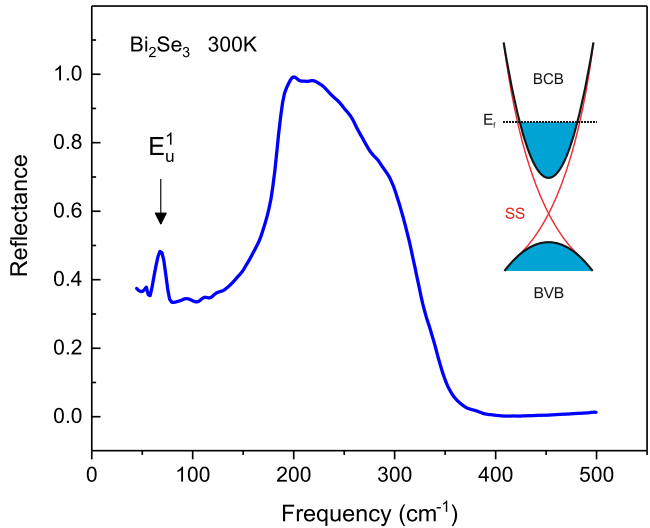}
\end{center}
\caption{\label{fig1} Reflectance spectrum of the Bi$_2$Se$_3$ film measured at room temperature. The arrow marks the position of the line that corresponds to the $E_u^1$ phonon mode. The inset shows a simplified model of the band structure of Bi$_2$Se$_3$ in the vicinity of the bulk band gap. Here $E_f$ denotes the Fermi level, while BVB, BCB, and SS --- bulk valence band, bulk conduction band, and surface states, respectively.}
\end{figure}

The reflectance spectrum of binary Bi$_2$Se$_3$ in the terahertz range measured at 300 K using the Fourier spectrometer is shown in Fig. 1. The broad spectral band located from 150 to 350 cm$^{-1}$ is due to phonons of the BaF$_2$ substrate (the so-called Reststrahlen band). The peak near 70 cm$^{-1}$ corresponds to the bulk infrared-active $E_u^1$ phonon mode \cite{Richter}. Figure 2 illustrates the evolution of the shape of this line with temperature for binary and strontium-doped Bi$_2$Se$_3$. It can be seen immediately that in the case of Bi$_2$Se$_3$ its asymmetry is very low at room temperature and almost absent at 5 K, in contrast to earlier works on bulk Bi$_2$Se$_3$ crystals, in which a considerable asymmetry was observed in the whole range of temperatures (see e. g. \cite{LaForge, Mallett}). It should be noted that at room temperature the shape of the $E_u^1$ phonon line is rather insensitive to doping by strontium atoms (Fig. 2(a)). The profiles observed for Bi$_2$Se$_3$ and Bi$_{2-x}$Sr$_x$Se$_3$ are similar with a small difference in line position and width. As we show below, the spectral parameters of the Raman phonon lines are even less different for binary and strontium doped crystals. Thus, we can suppose that dopant strontium atoms do not distort the lattice significantly, affecting mostly the electron subsystem of the host crystal.

\begin{figure}
\begin{center}
\includegraphics{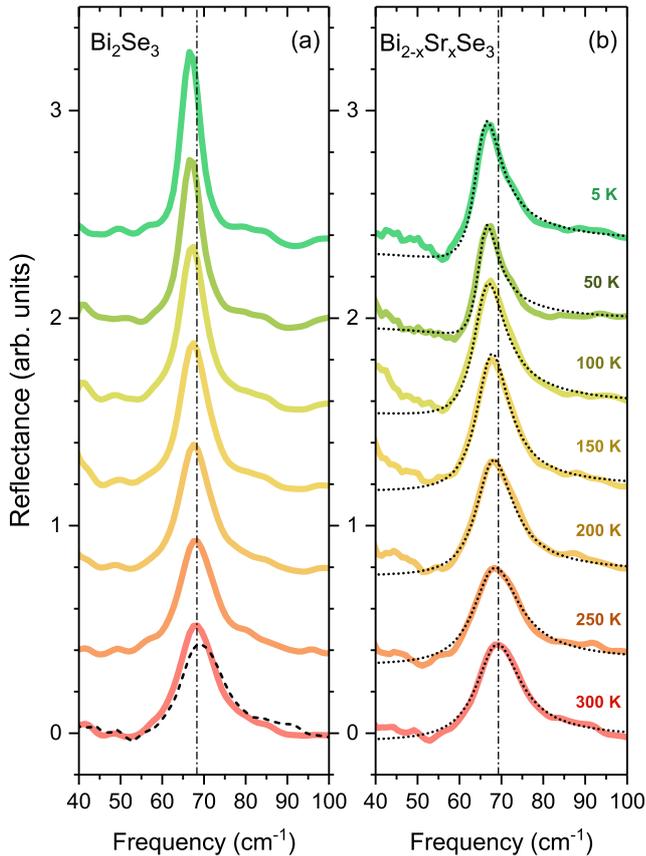}
\end{center}
\caption{\label{fig2} Parts of the reflectance spectra of Bi$_2$Se$_3$ (a) and Bi$_{2-x}$Sr$_x$Se$_3$ (b) in the vicinity of the $E_u^1$ phonon line measured at several temperatures from 5 to 300 K. The spectra are vertically shifted for clarity. Room temperature reflectance spectrum of Bi$_{2-x}$Sr$_x$Se$_3$ is duplicated in (a) as a dashed line for comparison. Dotted lines in panel (b) are fit curves obtained using Eq. 1.}
\end{figure}

A large difference between the spectra of Bi$_2$Se$_3$ and Bi$_{2-x}$Sr$_x$Se$_3$ occurs at lower temperatures $T\lesssim100$ K as the strong asymmetry of the $E_u^1$ phonon line of Bi$_{2-x}$Sr$_x$Se$_3$ develops (Fig. 2(b)). In order to better visualize the observed variation of the $E_u^1$ phonon line shape we performed fitting of the measured spectra in the vicinity of the line by the function
\begin{eqnarray}
R = R_0 - \frac{A\left(q+\frac{\nu - \nu_0}{\Gamma/2}\right)^2}{1 + \left(\frac{\nu - \nu_0}{\Gamma/2}\right)^2},
\end{eqnarray}
where $\nu_0$ is the resonance frequency, linewidth $\Gamma$ is associated with the phonon decay rate, while $q$ is the asymmetry parameter. This formula is written by analogy with the relation that describes Breit-Wigner-Fano resonances in order to quantify the asymmetry of the observed phonon line \cite{Fano}. The obtained temperature dependences of $\nu_0$, $\Gamma$, and $q$ are shown in Fig. 3. 

\begin{figure}
\begin{center}
\includegraphics{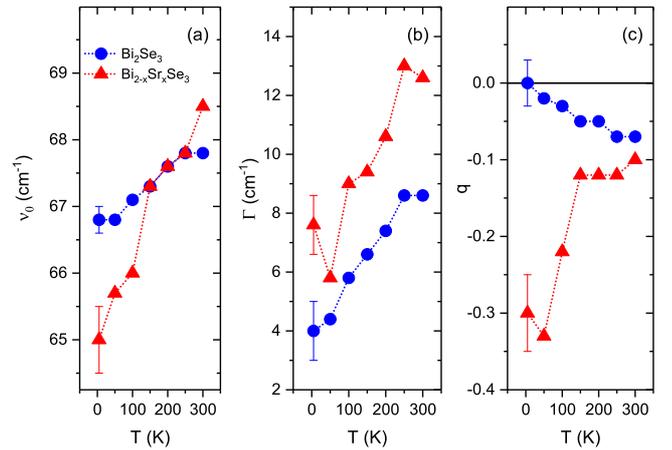}
\end{center}
\caption{\label{fig3} Dependences of the parameters of the $E_u^1$ spectral line extracted from the measured reflectance spectra. (a) -- Resonance frequency $\nu_0$. (b) -- Linewidth $\Gamma$. (c) -- Asymmetry parameter $q$. Data points for Bi$_2$Se$_3$ and Bi$_{2-x}$Sr$_x$Se$_3$ are shown as circles and triangles, respectively. The values of error illustrated by the error bars are related to all points of the corresponding set.}
\end{figure}

As can be seen from Fig. 3(b), the width of the $E_u^1$ line decreases with temperature both for Bi$_2$Se$_3$ and Bi$_{2-x}$Sr$_x$Se$_3$, while being higher for the strontium doped film for every value of the temperature. Such behavior of $\Gamma$ is typical for phonons that experience anharmonic decay and is demonstrated by other optical modes of binary and doped Bi$_2$Se$_3$ \cite{Li, Irfan, Kim}. The obtained temperature dependence of the $E_u^1$ resonance frequency is, however, rather uncommon. Indeed, as follows from Fig. 3(a), $\nu_0$ decreases with temperature. This softening of the $E_u^1$ mode is in contrast to the dynamics of Raman-active phonon modes, which demonstrate hardening upon cooling the crystal \cite{Li, Irfan, Kim}. Such unusual behavior was observed earlier in Bi$_2$Se$_3$ (see e. g. \cite{LaForge, Dordevic1}) and in a related topological insulator Sn-BSTS \cite{Li2}. In the latter case it was suggested that such softening could be the result of coupling between bulk $E_u^1$ phonons and surface electronic states, however the origin of this effect and the degree of its universality are still not clear. According to our data the decrease of $\nu_0$ upon cooling is faster for Bi$_{2-x}$Sr$_x$Se$_3$ than for Bi$_2$Se$_3$, resulting in the overall frequency shift $\Delta\nu_0\sim$ 4 cm$^{-1}$ compared to $\Delta\nu_0\sim$ 1 cm$^{-1}$ for the undoped crystal. Thus, the addition of strontium atoms enhances the softening of the $E_u^1$ phonon mode at low temperatures.

The most profound effect, however, the strontium doping has on the asymmetry of the $E_u^1$ spectral line profile. As can be seen from Fig. 2(a) and Fig. 3(c), for the binary Bi$_2$Se$_3$ a certain small asymmetry observed at 300 K tends to zero as the temperature is lowered to 5 K. On the contrary, for Bi$_{2-x}$Sr$_x$Se$_3$ the asymmetry parameter $q$ decreases considerably at temperatures below $\sim$ 150 K, while the concomitant almost three-fold increase of the absolute value of $q$ results in a pronounced Fano-like shape of the $E_u^1$ line in the spectra measured at 100, 50, and 5 K (see Fig. 2(b)).

We suggest the following interpretation of these observations. The Fano effect detected in our experiments for the infrared-active $E_u^1$ phonon mode implies the interaction of this mode with a continuum of states, the optical transitions to which are allowed in the dipole approximation. The characteristic temperature, near which the considerable asymmetry of the $E_u^1$ line develops, can be roughly estimated as  $\sim$ 100 K.  This value corresponds to the energy of $kT\approx$ 70 cm$^{-1}$ and is therefore very close to the frequency of the $E_u^1$ phonon mode. Such behavior can be accounted for if we assume that the optical transitions, which interfere with the absorption of the $E_u^1$ phonon, occur from surface to bulk electronic states that lie in the vicinity of the Fermi level and are separated by the energy of this phonon $\hbar\omega_\mathrm{ph}\sim$ 70 cm$^{-1}$. In electron-doped Bi$_2$Se$_3$ crystals (as in Bi$_{2-x}$Sr$_x$Se$_3$) the Fermi level is located relatively high in the conduction band ($\gtrsim$ 0.2 eV above its minimum), and in this energy region the surface states approach the conduction band rather closely, (see the inset to Fig. 1) so that the joint density of states at $\omega = \omega_\mathrm{ph}$ can be sufficiently large. At higher temperatures the Fermi distribution is smeared leading to the population of states above the Fermi level and to Pauli blocking of the transitions to these states \cite{Tang, Xu}. At the same time the occupation of the states below the Fermi level decreases, further reducing the probability of the corresponding optical transitions. As a result, the observed asymmetry of the phonon spectral line decreases. Lowering the temperature reverses this trend and the pronounced Fano effect appears at $T\lesssim\hbar\omega_\mathrm{ph}/k$. 

If we use this interpretation, then the effect of strontium doping has a straightforward explanation. For relatively thick epitaxial films of Bi$_2$Se$_3$ the overlap of surface and bulk electron wave functions is expected to be rather small, and so should be the probability of the optical transitions between these states. Moreover, the $E_u^1$ phonon that is probed by reflectance measurements is the bulk phonon, since the optical thickness of the epitaxial films that we studied in the experiments is moderate in the terahertz range. Thus, the observed Fano effect for the $E_u^1$ line should be small or absent. Doping a Bi$_2$Se$_3$ crystal with atoms lighter than bismuth decreases spin-orbit interaction and increases penetration of wave functions of the surface electronic states into the bulk. Recently this property was discussed as an explanation of the phonon Fano line shape in the indium-doped Bi$_2$Se$_3$ \cite{Sim}. A strontium atom is even lighter than indium, and the effect of Sr-doping on the overlap of the surface and bulk electronic wave functions should be rather large, leading to a considerable Fano effect. Exactly this difference between binary and Sr-doped Bi$_2$Se$_3$ was observed in our experiments, as the asymmetry of the $E_u^1$ line in Bi$_2$Se$_3$ was small at 300 K and decreased upon cooling, while for Bi$_{2-x}$Sr$_x$Se$_3$ the Fano effect was pronounced, demonstrating the specific temperature dependence in agreement with the simple model presented above.

Enhanced interaction of a phonon mode with electrons can result in an increase of the linewidth of the corresponding spectral line. However, the character of temperature dependences of $\Gamma$ obtained in our experiments for Bi$_2$Se$_3$ and Bi$_{2-x}$Sr$_x$Se$_3$ is similar (see Fig. 3(b)). The damping of the $E_u^1$ phonon is higher on average for Bi$_{2-x}$Sr$_x$Se$_3$, but there are no considerable peculiarities at low temperatures (taking into account experimental error). The fact that $\Gamma(T)$ curves for Bi$_2$Se$_3$ and Bi$_{2-x}$Sr$_x$Se$_3$ do not demonstrate such contrasting behavior as the $q(T)$ dependences can be due to the rather low maximal variation of the asymmetry parameter obtained in our case ($|q|\ll1$). For example, in the studies of the phonon Fano effect in TaAs and graphene similar or lower changes of $\Gamma$ were observed at much larger values of $q\gtrsim1$ \cite{Xu, Tang}. Here we can suppose that the value of $\Gamma$ is determined mostly by the anharmonic decay of the $E_u^1$ phonon, and the small influence of the electron-phonon interaction is difficult to notice on this background regarding the achieved accuracy. 

Electron-phonon interaction can also be the origin of phonon renormalization and of the corresponding frequency shift. As can be seen from Fig. 3(a), the softening of the $E_u^1$ phonon mode is stronger at lower temperatures for Bi$_{2-x}$Sr$_x$Se$_3$. As was mentioned above, the doping by strontium increases penetration of the surface electronic wave functions into the bulk, which can lead to the enhanced interaction of the latter with bulk phonons. However, the fact that the $E_u^1$ phonon mode softens upon cooling already in binary Bi$_2$Se$_3$ makes it difficult to distinguish interaction with bulk and surface electrons as possible contributions to the $E_u^1$ phonon frequency shift. 

One more probable explanation of the anomalous softening of the $E_u^1$ phonon mode could be its anharmonic interaction with vibrational modes of the crystal that are localized at impurities or defects. If the frequencies of these localized phonons were located below the $E_u^1$ mode, then this interaction would shift the $E_u^1$ mode to higher frequencies. If these lines were broad enough, an additional effect would be a small asymmetry of the $E_u^1$ line (can be illustrated as a classical analog of the Fano effect \cite{Joe}). Since the anharmonic phonon interactions weakens upon lowering the temperature, this model is consistent both with the gradual decrease of $q$ and $\nu_0$ observed for Bi$_2$Se$_3$. However, having in view the high crystalline perfection of the studied epitaxial films, this interpretation seems to be less likely. 

\begin{figure}
\begin{center}
\includegraphics{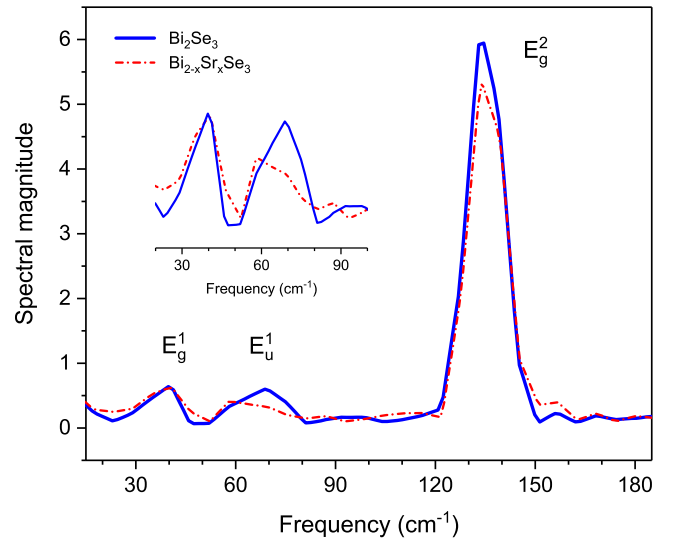}
\end{center}
\caption{\label{fig4} Phonon spectra of Bi$_2$Se$_3$ (solid line) and Bi$_{2-x}$Sr$_x$Se$_3$ (dash-dotted line) obtained by applying Fourier transformation to the time-resolved response of the samples to the intense terahertz pulse. The inset shows magnified parts of the spectra containing $E_g^1$ and $E_u^1$ lines.}
\end{figure}

The discussion above was focused on the dynamics of the bulk $E_u^1$ phonon mode, since reflectance measurements effectively probe the whole volume of the thin epitaxial films. However, we managed to study the effect of strontium doping on the surface counterpart of the $E_u^1$ phonon mode using coherent resonant excitation of these lattice vibrations by terahertz pulses as described above. The obtained phonon spectra are shown in Fig. 4. Here one can see three lines, two of which at 40 cm$^{-1}$ and 135 cm$^{-1}$ correspond to $E_g^1$ and $E_g^2$ bulk Raman active phonon modes, respectively. 
As was mentioned above, it is essentially the anisotropy of transmittance that is measured in such experimental configuration. However, due to symmetry restrictions only the surface counterpart of the bulk $E_u^1$ phonon mode can modulate the anisotropy of the refractive index and appear in the calculated spectra. The symmetry of the surface of a Bi$_2$Se$_3$ crystal is $C_{3v}$ unlike $D_{3d}$ symmetry of the bulk, which leads to a different classification of the phonon modes. Here we nevertheless refer to this surface mode as $E_u^1$ for clarity. 

It follows from the spectra in Fig. 4 that neither the resonance frequencies nor the line shapes of the $E_g^1$ and $E_g^2$ bulk Raman modes change noticeably upon doping of Bi$_2$Se$_3$ by strontium atoms, which implies that the concomitant distortion of the crystal lattice is very small. The surface $E_u^1$ mode, however, softens already at room temperature, as the corresponding spectral line shifts to lower frequencies. The absolute value of the shift can be roughly estimated as $\sim$ 3 cm$^{-1}$, which is much larger than the $\lesssim$ 1 cm$^{-1}$ small frequency shift observed for the bulk $E_u^1$ phonon mode of Bi$_{2-x}$Sr$_x$Se$_3$ in the reflectance spectrum measured at room temperature (see Fig. 2(a) and Fig. 3 (a)). Therefore, we can conclude that the interaction of the surface $E_u^1$ phonon mode with surface electrons is considerably larger than for its bulk counterpart. 

In summary, we observed anomalous behavior of the $E_u^1$ phonon mode in epitaxial single-crystal Bi$_{2-x}$Sr$_x$Se$_3$ films, which consisted in a Fano-like shape of the corresponding spectral line in the reflectance spectra and enhanced low temperature softening. In order to interpret this result, we considered it to be the combined effect of electron-phonon interaction and interference of lattice absorption with optical transitions between surface and bulk electronic states. Applying a specific surface sensitive method of ultrafast terahertz spectroscopy we obtained evidence of even stronger electron-phonon interaction at the surface of the Bi$_{2-x}$Sr$_x$Se$_3$ crystal. 

The reported study was funded by the Russian Foundation for Basic Research, project number 20-02-00989.

\end{document}